\documentclass[ASNA,twocolumn]{USG} 
\usepackage{anyfontsize} %

\usepackage{colortbl}   
\usepackage{xcolor}     

\usepackage{steinmetz}
\graphicspath{{./images/}}
\articletype{ORIGINAL ARTICLE}%

\received{}
\revised{}
\accepted{}
\journal{}
\volume{0}
\copyyear{2026}
\startpage{1}
\articledoi{10.1002/0000}

\begin{document}
\title{Mie-Resonant Dielectric Waveguides for Purcell-Enhanced Exciton Emission in Layered InSe}
\author[1]{A. I. Veretennikov}
\author[2]{E.A. Shepelev}
\author[1]{P.A. Alekseev}
\author[1]{I.A. Eliseyev}
\author[2]{A.S. Shorokhov}
\author[2]{A.A. Fedyanin}
\author[1]{M.V. Rakhlin}

\titlemark{Mie-Resonant Dielectric Waveguides for Purcell-Enhanced Exciton Emission in Layered InSe}

\address[1]{\orgname{Ioffe Institute, }%
\orgaddress{\state{194021 St.~Petersburg, }\country{Russia}}}

\address[2]{\orgdiv{Faculty of Physics, }\orgname{Lomonosov Moscow State University, }%
\orgaddress{\state{119991 Moscow, }\country{Russia}}}

\corres{M.V. Rakhlin  (\email{maximrakhlin@mail.ru})}






\abstract[ABSTRACT]{Layered van der Waals semiconductors are promising active materials for nanoscale photonic and optoelectronic devices because their excitonic emission can be integrated with heterogeneous photonic architectures. For on-chip applications, this emission must be efficiently coupled to guided modes, while its recombination dynamics should be controlled by the local photonic environment. Although dielectric waveguides enable such integration, substantial control over radiative recombination generally requires resonant engineering of the local photonic density of states. Resonant dielectric nanostructures provide such control by modifying the photonic environment while preserving compatibility with guided-wave photonic architectures. Here, we demonstrate Purcell-enhanced excitonic emission from a thin InSe flake integrated with a Mie-resonant Si$_3$N$_4$ waveguide. The structure incorporates a resonant nanoparticle array with a resonance that overlaps the InSe PL band, thereby enhancing excitonic coupling to the guided mode. Optical spectroscopy confirms the designed resonance, while micro-photoluminescence measurements reveal enhanced and spectrally selective waveguide-coupled emission. Time-resolved photoluminescence measurements show a threefold shortening of the excitonic decay time relative to planar InSe. Analysis of the decay dynamics using a simple rate-equation model yields an effective Purcell factor of approximately 3 for the dominant out-of-plane excitonic emission channel. These results establish Mie-resonant dielectric waveguides as a compact platform for on-chip control of excitonic recombination in layered semiconductors.
}







\maketitle


\section{Introduction}\label{intro}
Atomically thin van der Waals (vdW) semiconductors have emerged as promising active materials for nanoscale photonic and optoelectronic devices owing to their robust excitonic emission, thickness-dependent optical properties, and compatibility with heterogeneous integration \cite{wilson2021excitons,mueller2018exciton}. Their van der Waals nature enables transfer-based assembly on pre-fabricated photonic chips, offering a flexible route toward hybrid nanophotonic devices. For practical on-chip devices, however, the emitted light must be efficiently coupled to guided optical modes, while the excitonic recombination dynamics should be controlled by the local photonic environment. Achieving both requirements remains challenging because the coupling efficiency is strongly affected by the spatial and orientational overlap of excitonic dipoles and photonic modes \cite{lamprianidis2022directional,verhart2014single,peyskens2019integration}.

Among layered semiconductors, III--VI metal chalcogenides such as InSe are particularly attractive for waveguide-integrated nanophotonics. Few-layer InSe exhibits near-infrared photoluminescence (PL) with radiative lifetimes on the order of several nanoseconds  \cite{Venanzi2020,Borodin2024} and predominantly out-of-plane (OP) orientation of the luminescent excitonic dipoles \cite{Rybkovskiy2014,Brotons-Gisbert2019}. This dipole configuration favors coupling to dielectric waveguide modes with a strong OP electric-field component and allows excitonic recombination dynamics to be engineered via the local photonic density of states \cite{jun2009broadband,liebermeister2014tapered}. Nevertheless, previously reported waveguide-integrated 2D vdW emitters have mainly been optimized for coupling excitonic emission into guided modes, with typical coupling efficiencies limited to 5--15\% \cite{Tonndorf2017,Errando-Herranz2021}. In addition, conventional waveguide (CW) geometries usually provide limited control over the radiative recombination rate, leaving the emission dynamics largely governed by the intrinsic recombination channels of the semiconductor \cite{Wang2019times,tamalampudi2014}.

A promising strategy for overcoming these limitations is to integrate layered emitters with low-loss resonant dielectric nanostructures. By confining optical fields at the nanoscale, such structures can modify the local photonic density of states, strengthen exciton--mode coupling, and alter spontaneous emission rates through the Purcell effect \cite{krasnok2012all,Gartman2023,obydennov2024asymmetric}. Silicon nitride (Si$_3$N$_4$) is especially suitable for this purpose owing to its broad transparency window, low optical losses, and compatibility with standard nanophotonic fabrication techniques \cite{peyskens2019integration,Bauters2011,blumenthal2018silicon}. While resonant vdW--based waveguide structures have recently demonstrated enhanced coupling of excitonic emission to guided modes \cite{Rakhlin2025}, direct experimental evidence that such resonant waveguide (RW) structures can modify excitonic recombination dynamics has remained limited.

 \begin{figure}
    \centering
    \includegraphics[width=\linewidth]{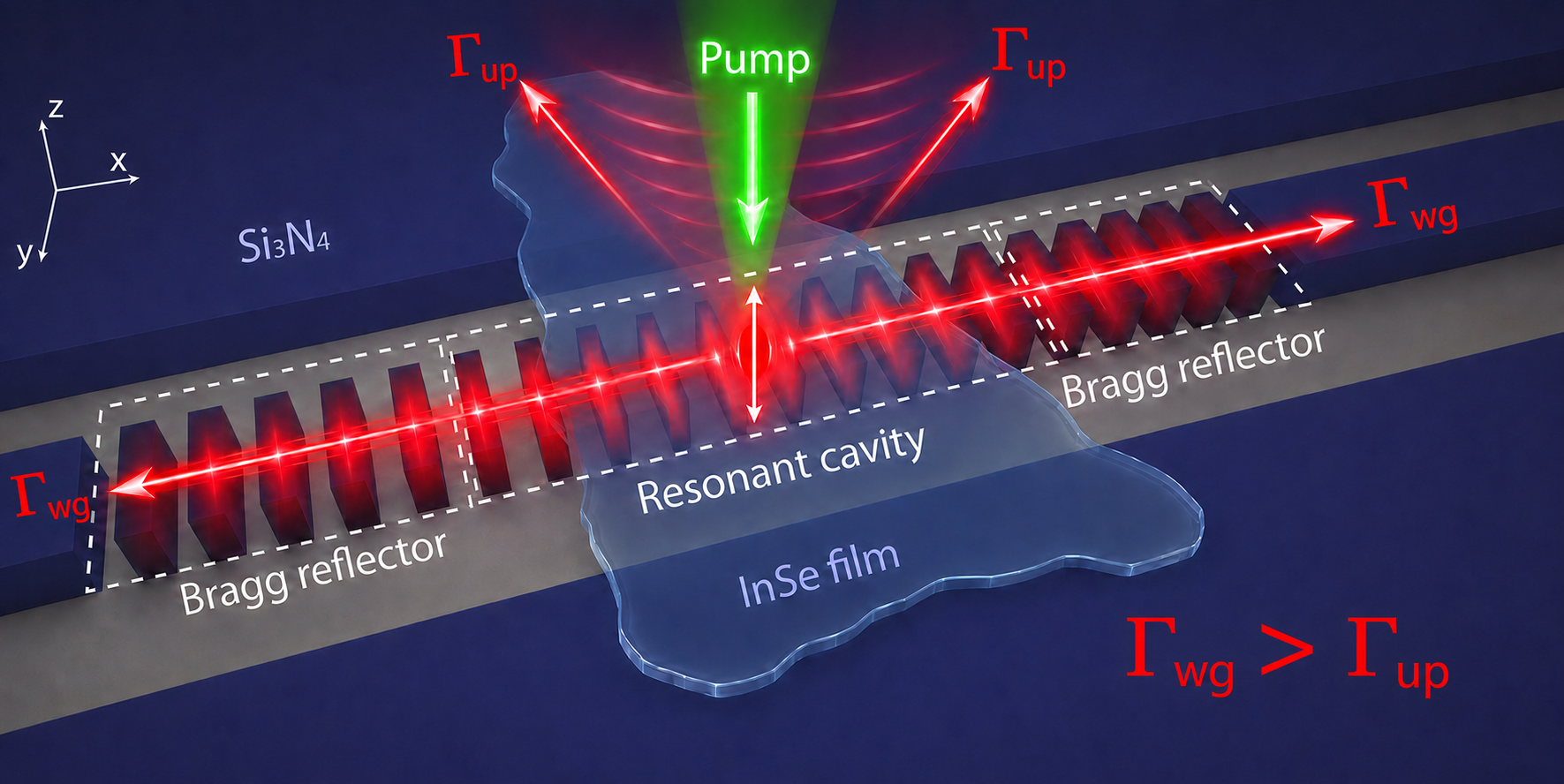}
    \caption{Schematic illustration of the optical coupling between the RW structure and the exciton emitter in the thin InSe film on top of it. \(\Gamma_\mathrm{wg}\) and \(\Gamma_\mathrm{up}\) schematically denote the effective emission-rate contributions associated with coupling to the guided resonant mode and radiation into the upper half-space, respectively.}
    \label{sample_scheme}
\end{figure}

Here, we demonstrate Purcell-enhanced excitonic emission from a thin InSe flake integrated with a Mie-resonant Si$_3$N$_4$ waveguide. The waveguide incorporates a resonant nanoparticle array, whose optical resonance is spectrally matched to the InSe PL band to enhance coupling of excitonic emission to the guided mode. Optical spectroscopy confirms the predefined resonance of the fabricated structure, whereas micro-photoluminescence ($\mu$-PL) measurements reveal enhanced and spectrally selective waveguide-coupled emission. Time-resolved photoluminescence decay curves show a reduction of the excitonic decay time relative to planar InSe, with the fast and middle components shortened by factors of 3 and 2.1, respectively. Analysis of the decay dynamics using a simple rate-equation model yields an effective Purcell factor of approximately 3 for the dominant OP-mediated guided-emission channel. These results establish Mie-resonant dielectric waveguides as a compact platform for on-chip control of excitonic recombination in layered semiconductors and for the development of active vdW nanophotonic components.

 \section{Design, Simulation and Fabrication of the Resonant Waveguides}\label{numcalc}

The investigated hybrid structure consists of a silicon nitride RW integrated with a 13-nm-thick InSe film, as schematically shown in Figure~\ref{sample_scheme}. The RW comprises two distributed Bragg reflectors and a central resonant cavity. The cavity is implemented as a chain of Mie-resonant dielectric nanoparticles whose thickness decreases from the edges toward the center of the cavity following a parabolic profile. The individual nanoparticles support localized Mie-type resonances, while their spatially varying geometry induces a parabolic distribution of the effective refractive index within the structure, thereby enabling efficient optical-field localization inside the resonator~\cite{Ahn2010}.

\begin{figure}[hbtp]
    \centering
    \includegraphics[width=\linewidth]{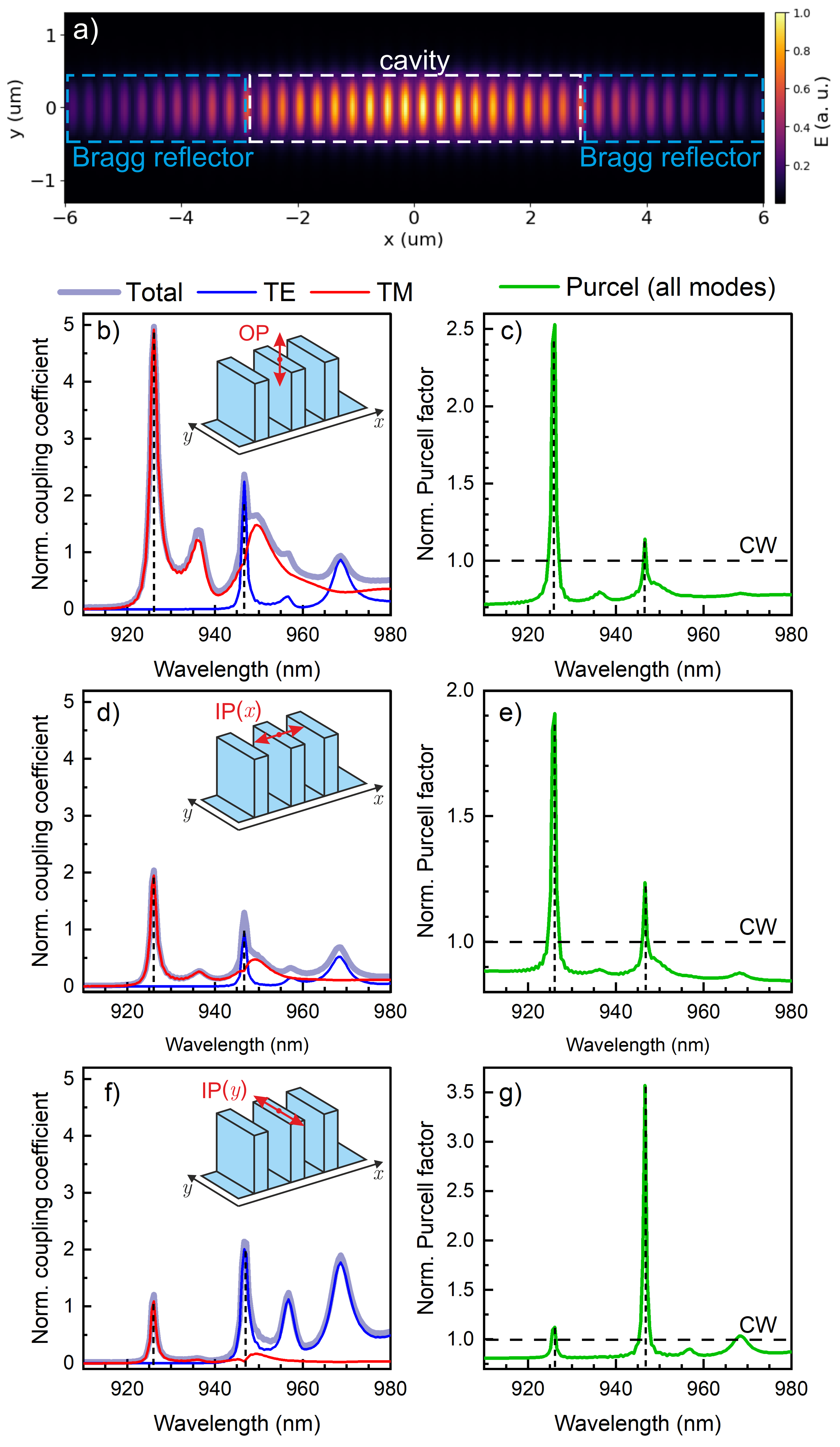}
    \caption{Numerical analysis of dipole coupling to the Mie-resonant Si$_3$N$_4$ waveguide. (a) Simulated electric-field distribution at the wavelength of the cavity resonance of 927 nm effectively coupled to the TM mode of the waveguide. (b,d,f) Coupling efficiency spectra calculated for different dipole orientations and normalized to the corresponding values for the conventional waveguide. Vertical dashed lines indicate the spectral positions of the most prominent resonances. Blue and red lines depict decomposition of the total coupling coefficient into TE and TM components. Insets show schematic representations of the nanoparticle geometry and the corresponding dipole orientation. (c,e,g) Effective Purcell factor spectra calculated for the same dipole orientations. }
    \label{Purcell}
\end{figure}

To identify the optimal geometrical parameters of the RW, numerical simulations were carried out using the finite-difference time-domain (FDTD) method \cite{Gartman2023}. The numerical model included not only the Si$_3$N$_4$ resonant waveguide but also the 13-nm-thick InSe flake placed on top of the resonant section, corresponding to the experimental geometry. Therefore, the effect of the finite InSe thickness and its dielectric response on the resonance wavelengths, coupling efficiencies, and Purcell factors was taken into account in the simulations. Based on the simulation results, the following waveguide parameters were adopted: waveguide width $W = 0.85$~$\mu$m, height $H = 0.4$~$\mu$m, nanoparticle period $\Lambda = 0.3$~$\mu$m, cavity length $L_{\text{cav}} = 6$ $\mu$m, and Bragg reflector length $L_{\text{ref}} = 3$ $\mu$m. The thickness of the nanoparticles within the cavity was varied from $l = 150$~nm in the center to $l = 170$~nm at the edges of the cavity. The calculated optical-field intensity distribution in the vicinity of the resonant cavity (Figure~\ref{Purcell}a) confirms the field localization at the resonant wavelength.


For the waveguide geometry described above, we calculated the Purcell factor and coupling efficiency for three characteristic dipole orientations (see the insets in Figures~\ref{Purcell}b, d, f): OP, IP along the waveguide ($x$-axis) and IP across the waveguide ($y$-axis). To account for the random position of the excited dipole relative to the center of the nanoparticle, we averaged all spectra over different $x$ and $y$ coordinates of the dipole in the region of the cavity (details on the averaging procedure are provided in the Supporting Information). The calculated spectra are shown in Figs.~\ref{Purcell}b--g. The resonant response of the structure originates from Mie-type modes supported by the dielectric nanoparticles and their hybridization with the guided mode of the waveguide. When the dipole is positioned at the nanoparticle center (insets in Fig. 2), an OP dipole excites only the TM mode in the strip part of the waveguide, while a $y$-polarized IP dipole excites only the TE mode. Displacing the dipole toward the nanoparticle edge increases coupling to the orthogonal waveguide polarization. Consequently, the averaged spectra contain contributions from both TE and TM polarizations. Different effective refractive indexes for the waveguiding modes lead to splitting of the main cavity resonance into TE and TM  components. Hence, the averaged coupling coefficient spectra contains two distinct resonant peaks near 927 and 947~nm corresponding to TM and TE polarized light wave in the strip-waveguide section of the RW, respectively. Full decomposition of the coupling coefficient on TE and TM components is presented in Figure~\ref{Purcell} with blue and red curves. As a result of spectrally sensitive resonant coupling, spectra of the Purcell factor also demonstrate two narrow features at the respective wavelengths (dashed vertical lines in Fig.~\ref{Purcell}).

Efficient excitation of the TM mode by the OP dipole results in a fivefold increase in coupling efficiency compared with the conventional waveguide of the same geometrical parameters. The strong spatial and spectral overlap with the resonant local field additionally leads to a Purcell factor of approximately 2.6. This combination makes the OP orientation the most favorable for coupling to the resonant waveguide mode. In contrast, the IP dipoles couple less efficiently to the TM mode. For these orientations, a high Purcell factor is achieved only for the \(y\)-polarized dipole near 947 nm. 
In addition to their weaker coupling to the waveguide within the InSe PL band (900--1000 nm), IP dipoles are randomly oriented, and only a fraction of optically active IP excitons are \(y\)-polarized. Moreover, previous experiments on thin flakes of III-chalcogenides indicate that the contribution of IP dipoles to the overall PL is approximately five times lower than that of OP dipoles \cite{Xiao2020}. Consequently, the waveguide-coupled luminescence is expected to be dominated by OP-dipole emission, with the contribution of enhanced IP-dipole emission being at least an order of magnitude smaller.


\begin{figure}[t]
    \centering
    \includegraphics[width=\linewidth]{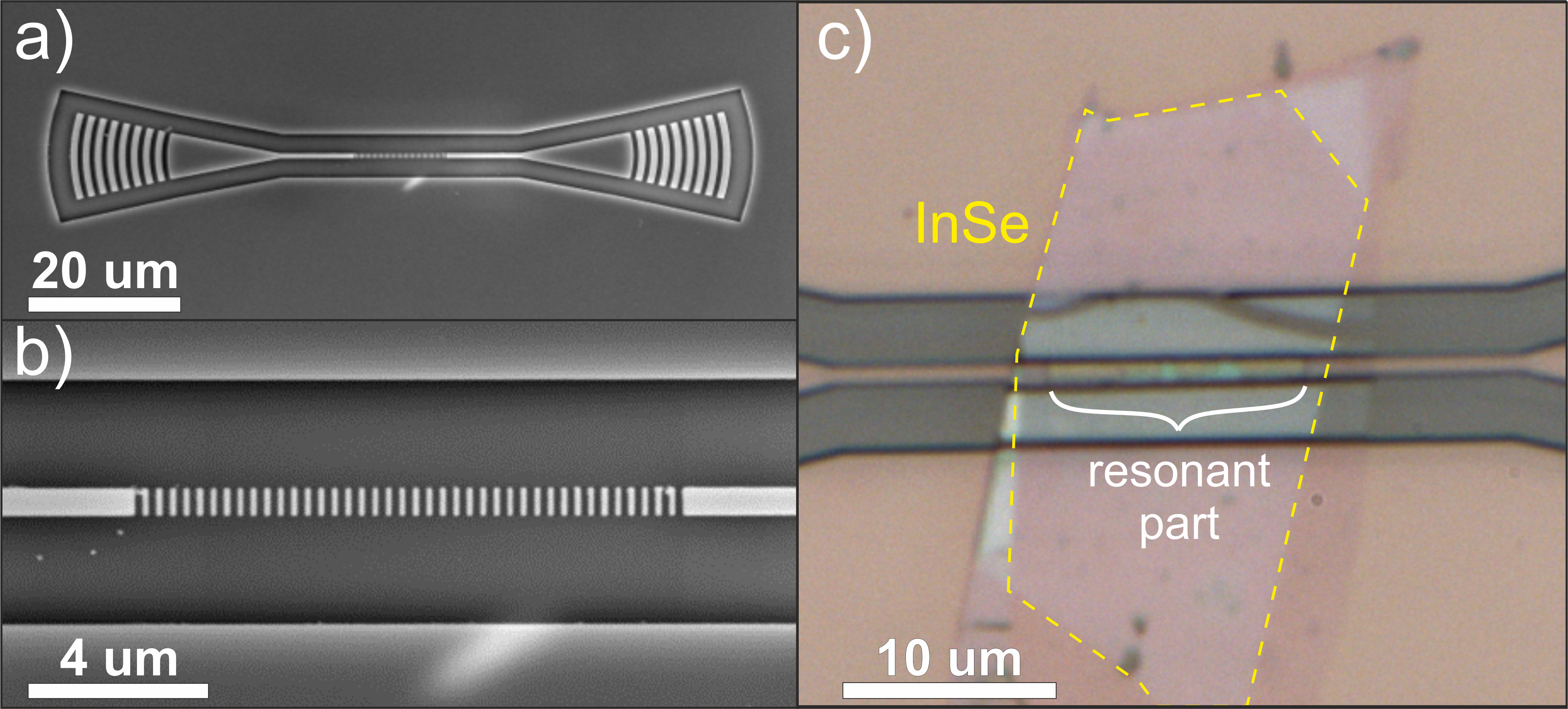}
    \caption{(a) Scanning electron microscope (SEM) image of the RW before the InSe film transfer. (b) Scaled SEM image of the resonant section of the RW. (c) Optical image of the flake transferred on the RW.}
    \label{SEM_img}
\end{figure}

The designed hybrid vdW--dielectric RW structure was fabricated to experimentally characterize its optical response. The Si$_3$N$_4$ waveguides were patterned using electron-beam lithography followed by reactive ion etching, while thin InSe flakes were prepared by mechanical exfoliation and transferred onto the photonic structures (see Appendix A.1 for fabrication details). Scanning electron microscopy images of the fabricated RW and its active section are shown in Figures~\ref{SEM_img}a,b, respectively. An optical image of the transferred InSe flake positioned over the resonant section is shown in Figure~\ref{SEM_img}c. Optical transmission measurements of the etched RW structures confirmed the presence of transmission peaks at predefined spectral positions (see Appendix~\ref{charact} for details). Additional characterization of the transferred flakes, including atomic force microscopy and micro-Raman mapping, is provided in the Supporting Information.



\begin{figure*}[htbp]
    \centering
    \includegraphics[width=0.85\linewidth]{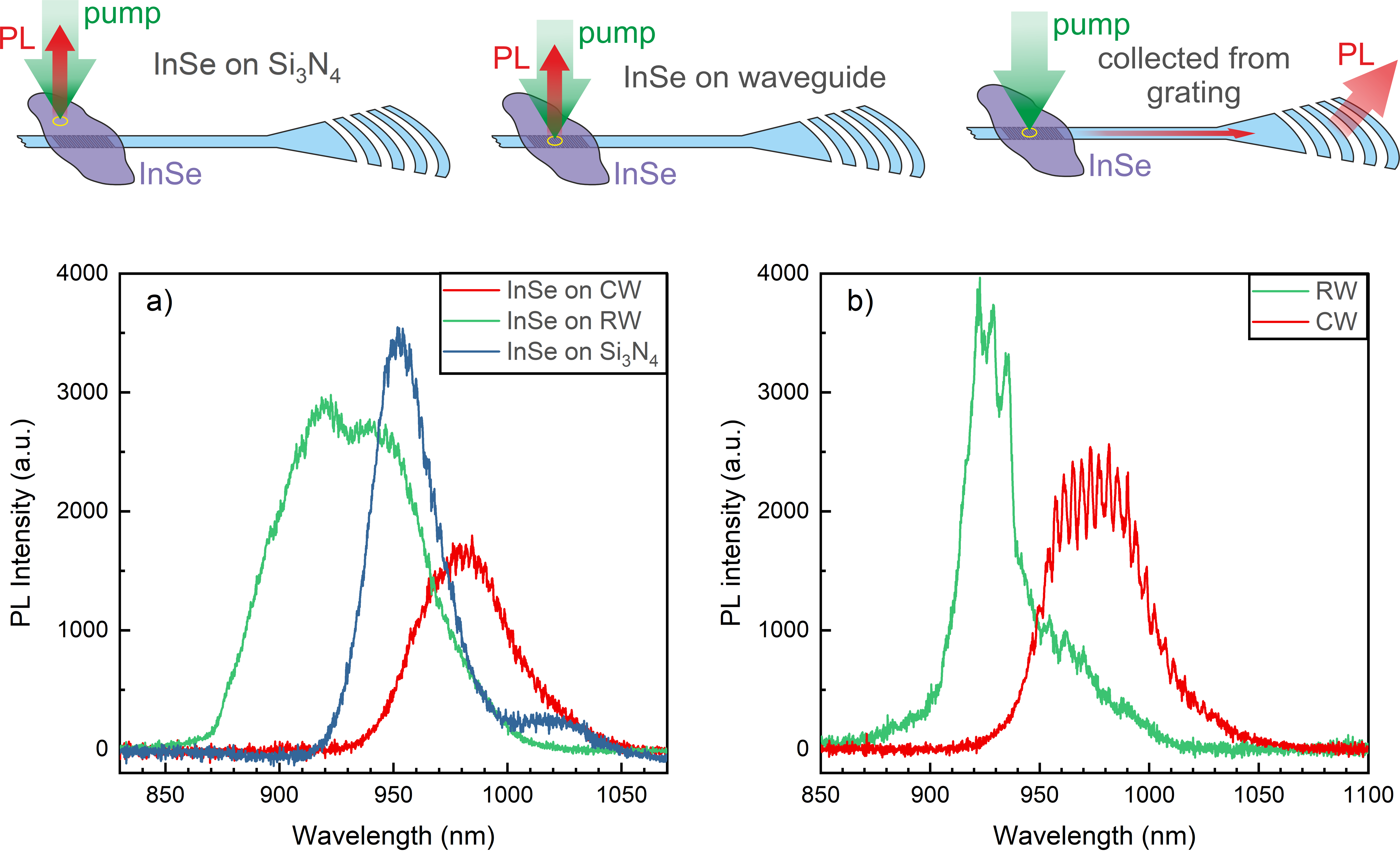}
    \caption{Results of \(\mu\)-PL measurements. The sketches above the graphs show the corresponding excitation and collection geometries. (a) \(\mu\)-PL spectra of an InSe flake on a planar substrate (blue curve), on the CW (red curve), and on the RW (green curve), measured in the back-scattering configuration. 
(b) \(\mu\)-PL spectra of the luminescence excited at the center of the waveguide and collected from the output grating for the CW (red curve) and RW (green curve) structures.}
    \label{PL_CW/RW}
\end{figure*}

\section{Enhanced Coupling of InSe Emission to the Guided Mode}

Figure~\ref{PL_CW/RW}a presents the $\mu$-PL spectra measured at different positions on the sample in the back-scattering configuration, as illustrated by the left and central sketches at the top of the figure. The integrated PL intensity collected from the center of the CW region (red curve) is approximately 30\% lower than that measured from an InSe flake on a planar substrate (blue curve). This decrease is consistent with partial redistribution of the emitted light into the guided mode of the CW, which was previously estimated to reach approximately 20\% in similar structures~\cite{Rakhlin2025}. In this geometry, part of the emitted radiation is redirected into the waveguide and is therefore not collected by the objective.

In contrast, the PL signal from the active region of the RW structure (green curve) is substantially enhanced relative to the planar reference. This enhancement can be attributed to resonant electromagnetic-field concentration and redistribution in the vicinity of the InSe flake induced by the Mie-type nanoparticles; similar effects have been reported for 2D semiconductors integrated with all-dielectric resonant nanostructures~\cite{Bucher2019,Ma2024,Deng2026}. In addition, the calculated coupling-efficiency spectra suggest that, in certain spectral regions, the normalized coupling efficiency of OP-dipole emission to the guided mode can be lower than that of the CW. In the back-scattering configuration, this reduced transfer into the guided channel can increase the fraction of emission collected by the objective. This behavior does not contradict the enhanced waveguide-coupled emission discussed below, because the two measurements probe different collection channels and different spectral components of the resonant response. The spectral modulation observed for both the RW and CW structures compared with planar InSe is attributed to imperfect spectral overlap between the InSe emission band and the wavelength-dependent coupling resonances of the waveguides.

To assess the effect of the resonant nanoparticles on coupling to the guided mode, we used a modified detection scheme. In this configuration, PL was excited at the center of the waveguide, while the signal was collected exclusively from the diffraction grating (see the right-hand sketch in Figure~\ref{PL_CW/RW}). This arrangement selectively probes the fraction of luminescence coupled to the guided mode. The corresponding spectra are shown in Figure~\ref{PL_CW/RW}b. For the CW device, the spectrum exhibits an approximately Gaussian envelope determined by the broadband waveguide transmission, superimposed with a periodic modulation arising from Fabry--Perot-like interference induced by the reflectance from the diffraction gratings.

In contrast, the RW spectrum displays a stronger and spectrally narrower feature associated with the predefined resonances of the structure. Because this narrow spectral region overlaps with both calculated resonances, the emitting dipole orientation cannot be unambiguously identified from the $\mu$-PL spectra alone. Nevertheless, the contribution of IP dipoles is expected to remain small for the reasons discussed above. Overall, the intense and spectrally narrow peak observed for the RW structure indicates improved spectral selectivity and stronger coupling to the guided mode enabled by the resonant grating.

\section{Time-Resolved Photoluminescence and Effective Purcell Enhancement\label{TRSection}}

The pronounced optical resonance supported by the RW structure is expected to affect not only the coupling efficiency but also the recombination dynamics of excitons whose emission is coupled to the resonant mode. To clarify the role of the RW resonances in the recombination process, we performed time-resolved photoluminescence (TRPL) measurements and extracted the corresponding PL decay times. In all TRPL experiments, background signals not associated with InSe emission were suppressed by selecting the spectral range of 920--960 nm (see Section \ref{TRmes} of Appendix for technical details).

\begin{figure}[h]
    \centering
    \includegraphics[width=\linewidth]{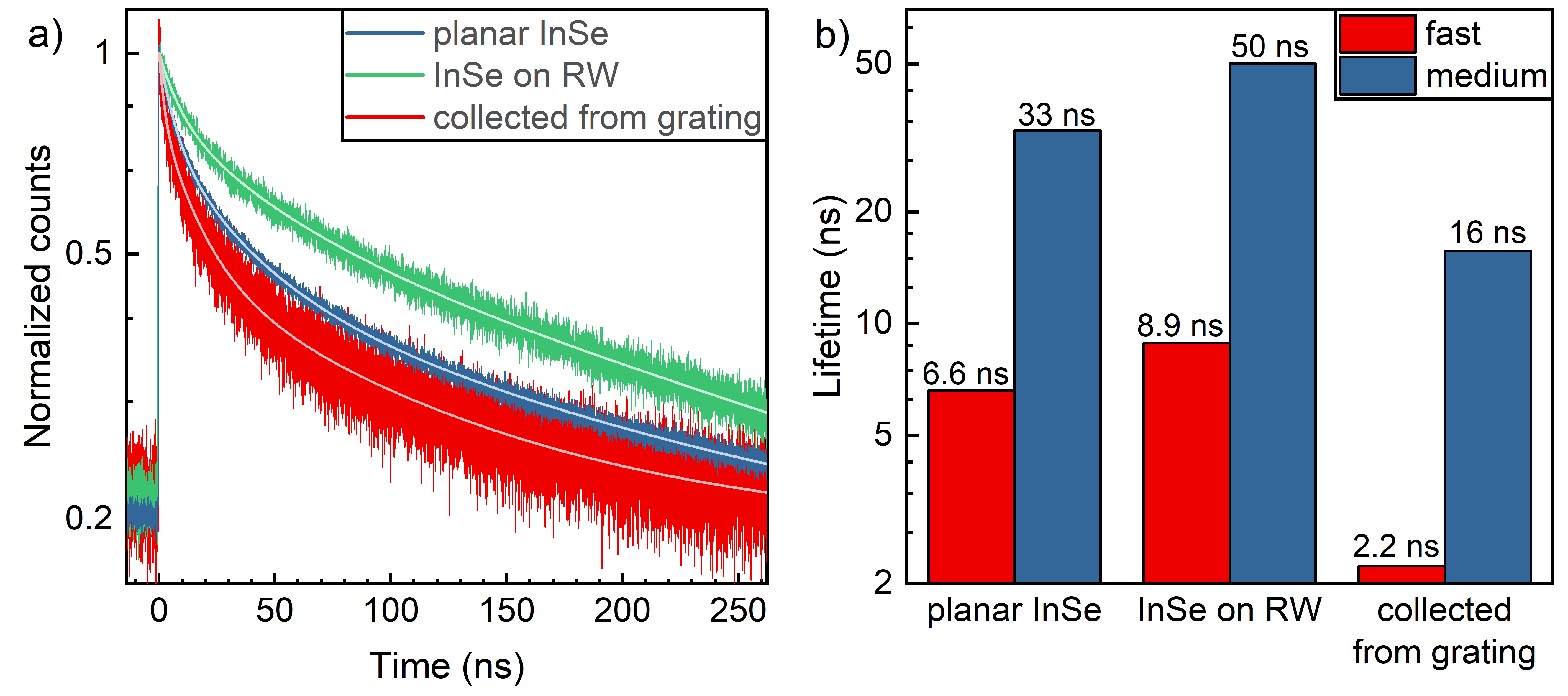}
    \caption{(a) PL decay curves on a logarithmic scale. The blue curve corresponds to PL from the InSe flake on a planar substrate, the green curve shows the PL decay measured at the center of the RW in the back-scattering configuration, and the red curve shows the PL decay of waveguide-coupled emission collected from the output grating. Solid lines represent exponential fits. (b) Recombination lifetimes extracted from the TRPL data in panel (a) on a logarithmic scale. Red and blue bars correspond to the fast and middle components, respectively.}
    \label{TRPL_data}
\end{figure}

Figure~\ref{TRPL_data}a shows normalized TRPL curves measured for an InSe flake on a planar substrate and for luminescence coupled from the flake onto the RW resonant part and collected from the output grating. The decay curve of the guided emission (red) exhibits a markedly faster PL decay than that of the planar reference (blue). The same trend is observed for all probed positions in the planar region and across the resonant area of the waveguide, supporting a resonator-induced origin of the observed lifetime reduction. Within this framework, the recombination rate is enhanced for the emission channel coupled to the resonant mode, whereas emission into weakly coupled channels may exhibit an effectively reduced rate due to the redistribution of the local density of optical states \cite{Bordo2012, Androvitsaneas2016}. This interpretation is supported by the TRPL signal measured from the RW region in the back-scattering geometry (green curve in Fig.~\ref{TRPL_data}a), which shows a slower decay than the planar reference. The same mechanism also explains the enhanced guided PL intensity observed for the RW structure, as discussed in the previous section. To verify that the observed lifetime modification is primarily associated with the resonant elements of the RW, we performed identical TRPL measurements on the CW structure using another InSe flake of similar thickness (see Section \ref{B4TRPL} of Appendix for TRPL data). No statistically significant changes in the PL decay times were observed for the CW structure, supporting the essential role of the resonator in the lifetime reduction.

Although the decay curves clearly indicate a reduction of the PL lifetime, a quantitative estimation of the Purcell factor requires identification of the independent recombination channels involved in the coupling process. In a recent study \cite{Borodin2024}, the PL decay of InSe was decomposed into "fast", "middle", and "slow" components associated with direct interband recombination, indirect phonon-assisted interband recombination, and defect-related recombination, respectively. Following this approach, we fitted all decay curves with a sum of three exponential functions, as described in Ref.~\cite{Borodin2024}; the resulting fits are shown as solid curves in Fig.~\ref{TRPL_data}a. For all TRPL curves, the lifetime of the "slow" component exceeded 150~ns, and this contribution was therefore excluded from the subsequent analysis. 

\begin{figure}[hbtp]
    \centering
    \includegraphics[width=\linewidth]{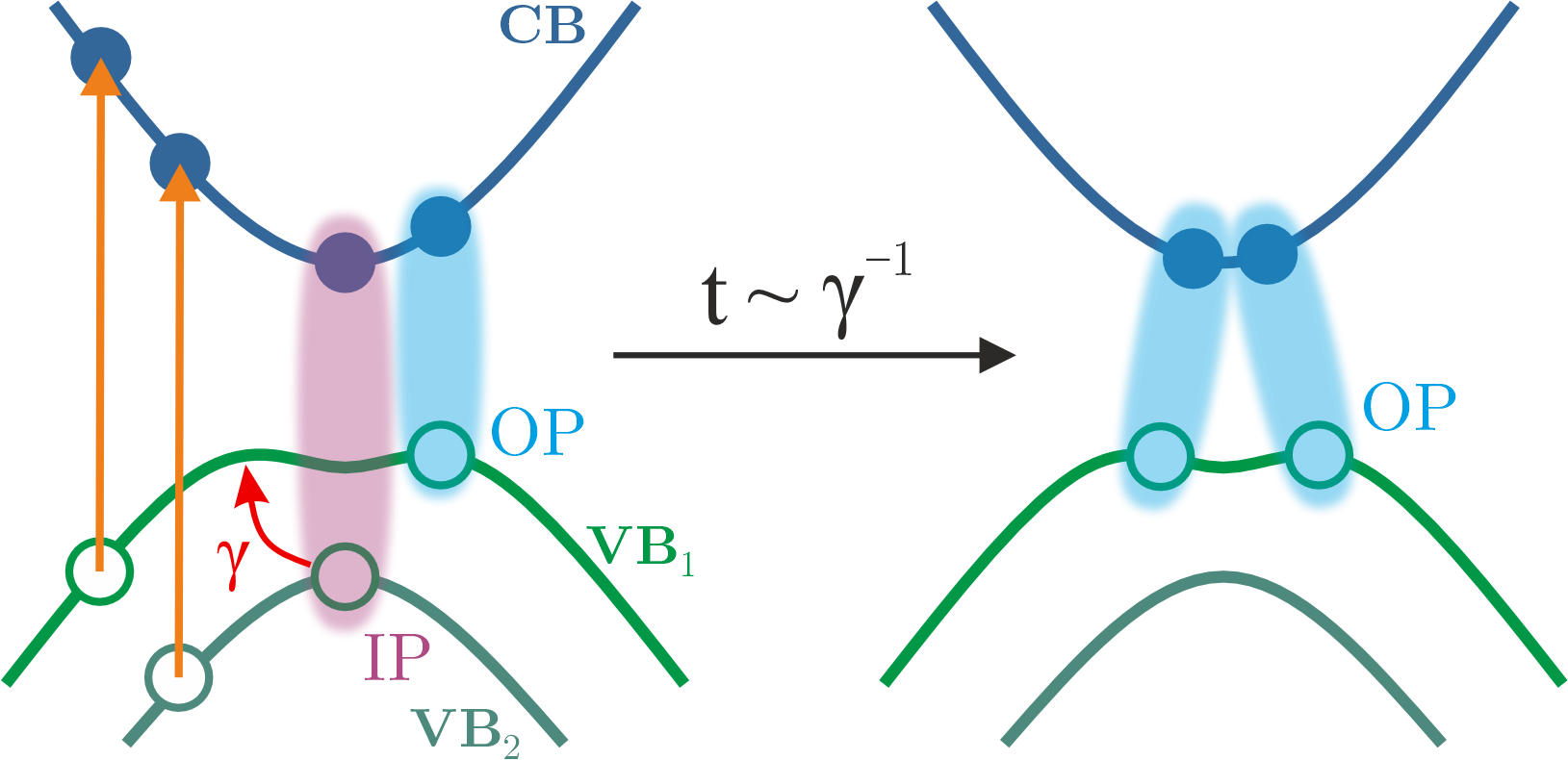}
    \caption{Illustration of exciton formation and interconversion in layered InSe. Pulsed optical excitation generates a mixed population of IP excitons (purple area) and OP excitons (blue area). Due to interband hole relaxation (red arrow), IP excitons can be converted into OP excitons with a characteristic timescale of approximately $\gamma^{-1}$. }
    \label{Pscheme}
\end{figure}

The extracted "fast" and "middle" lifetimes are summarized in Fig.~\ref{TRPL_data}b. We found the "fast" lifetime to decrease by a factor of 3, while the "middle" lifetime shortened only by 2.1.
This discrepancy can be eliminated using a rate-equation model that considers the generation, interconversion, and recombination dynamics of IP and OP excitons, as illustrated in Figure~\ref{Pscheme} and following the interpretation proposed in Ref. \cite{Borodin2024}. The pulsed optical excitation is assumed to generate a population of IP excitons, $N_{IP}$, through direct excitation of in-plane dipole-allowed transitions, while a population of OP excitons can also be generated due to strong mixing of valence-band states with $p_z$ and $p_x$–$p_y$ orbital character in layered InSe \cite{Rybkovskiy2014,Brotons-Gisbert2019}. In thin multilayer InSe flakes, the dense valence-subband structure can enable significant absorption even at 1.72 eV \cite{Zultak2020,Pasquale2022}. After excitation, holes relax toward the top of the valence band, effectively converting IP excitons into OP excitons at the transition rate $\gamma$. At the same time, both IP and OP excitons undergo radiative recombination with characteristic lifetimes $\tau_{IP}$ and $\tau_{OP}$, respectively. In the vicinity of the resonant cavity, the corresponding radiative decay rates are enhanced by the Purcell factors $F_p^{OP}$ and $F_p^{IP}$ for OP and IP excitons, respectively. To analyze the resulting kinetics, we use the following rate-equation model:
\begin{equation}\label{rateq}
    \begin{split}
        \frac{d}{dt}N_{IP}(t) =& -\frac{F_p^{IP}}{\tau_{IP}}N_{IP}(t)-\gamma N_{IP}(t),\\
    \frac{d}{dt}N_{OP}(t) =& -\frac{F_p^{OP}}{\tau_{OP}}N_{OP}(t)+\gamma N_{IP}(t),
    \end{split}
\end{equation}

with the initial conditions: \(N_{IP}(0)=\alpha,\ N_{OP}(0)=1-\alpha\). Here, $\alpha$ is the fraction of IP excitons relative to the total number of excited excitons. The exact solution for $N_{OP}(t)$ has the following two-exponential form:
\[N_{OP}(t)=A\exp\left(-\frac{F_p^{OP}}{\tau_{OP}}t\right)+B\exp\left(-\left(\frac{F_p^{IP}}{\tau_{IP}}+\gamma\right)t\right),\]
(see Appendix \ref{SIrateq} for analytical expressions for $A$ and $B$ coefficients). The first term describes direct OP exciton recombination and corresponds to the "fast" component, whereas the second term accounts for delayed refilling of the OP population from IP excitons and corresponds to the "middle" component. Both decay components are influenced by Purcell enhancement, but the resulting lifetime shortening is not identical. The fast component is shortened directly by the Purcell factor $F_p^{OP}$, whereas the middle component is shortened by an effective factor \(\tilde{F}_p=(F_p^{IP}+\gamma\tau_{IP})/(1+\gamma\tau_{IP})\). This reduced effective shortening arises because the middle component contains a relaxation-limited contribution from IP excitons, and the interband hole relaxation rate $\gamma$ remains unaffected by the resonant photonic environment. Using the experimentally measured value $\tau_{OP}=6.6$~ns, $\tau_{IP}/\tau_{OP}\approx11.4$ \cite{Gomes1993}, the experimentally extracted value  \((\gamma+\tau_{IP}^{-1})^{-1}=33\)~ns and $F_p^{IP}=3.2$, estimated from the weighted numerical Purcell factor and scaled by the ratio $F_{exp}^{OP}/F_{num}^{OP}$ (see Appendix \ref{SIrateq}), we obtain $\tilde{F}_p\approx2.1$ in good agreement with the experimentally observed shortening of the "middle" component. Thus, we estimate an effective Purcell factor $F_p^{OP}\approx3$ for the dominant OP-mediated decay channel, which is consistent with the numerical simulations and supports the dominant role of OP excitons in the observed lifetime shortening.

\section{Conclusion}

We have demonstrated resonator-induced control of excitonic emission dynamics in the hybrid vdW--dielectric nanophotonic structure consisting of the thin InSe flake integrated with the Mie-resonant Si$_3$N$_4$ waveguide. The waveguide incorporates a resonant nanoparticle array, engineered so that its optical resonance overlaps the InSe PL band and enhances coupling of excitonic emission to the guided mode. Optical transmission measurements confirmed the predefined resonance of the fabricated structure, while micro-photoluminescence measurements showed enhanced and spectrally selective waveguide-coupled emission from InSe in the resonant structure compared with the conventional strip waveguide.

Time-resolved photoluminescence measurements revealed an up to threefold shortening of the excitonic decay time relative to planar InSe. Control measurements performed on a conventional waveguide did not show a comparable lifetime reduction, supporting the resonator-induced origin of the observed lifetime shortening. By analyzing the decay kinetics with a rate-equation model that accounts for IP-to-OP exciton conversion, we estimated an effective Purcell factor of approximately 3 for the dominant OP-mediated guided excitonic luminescence.

These results provide experimental evidence that Mie-resonant dielectric waveguides can be used not only to enhance coupling of layered-semiconductor emission to guided optical modes, but also to modify excitonic recombination dynamics through resonant engineering of the local photonic environment. The demonstrated approach establishes hybrid InSe/Si$_3$N$_4$ resonant waveguides as a compact platform for on-chip control of exciton emission in layered semiconductors and for the development of active vdW nanophotonic components.

\bmsubsection*{Author Contributions}

A.I.V. prepared the InSe flakes, performed micro-photoluminescence measurements, and wrote the original draft. 
E.A.S. and A.S.S. performed numerical simulations and optical transmission measurements. 
P.A.A. performed AFM measurements. 
I.A.E. performed micro-Raman measurements. 
A.S.S., A.A.F., and M.V.R. contributed to conceptualization. 
A.I.V., A.S.S., A.A.F., and M.V.R. contributed to methodology. 
M.V.R. supervised the project and revised the manuscript. 
All authors discussed the results and approved the final version of the manuscript.

\bmsubsection*{Acknowledgments}
The work of M.V.R. was supported by a grant from the Russian Science Foundation (no. 24-72-00148, https://rscf.ru/project/24-72-00148/). The work was carried out using the equipment purchased partly by Moscow State University Development Program. Numerical modeling and nanostructure fabrication was conducted under the state assignment of Lomonosov Moscow State University. 

\bmsubsection*{Financial Disclosure}

The authors declare no financial interests.

\bmsubsection*{Conflicts of Interest}

The authors declare no conflicts of interest.

\bibliography{wileyNJD-Chicago}

\bmsubsection*{Supporting Information}

Additional supporting information can be found online in the Supporting Information section.\\
\textbf{Supporting File}:

\appendix

\bmsection{Samples and fabrication\label{samples}}
\bmsubsection{Fabrication of the Si$_3$N$_4$ structure\label{etch}}
\vspace*{12pt}

Experimental samples based on a silicon nitride waveguide system were fabricated using the following methods:
\begin{enumerate}
\item A silicon wafer was placed in a furnace at high temperature in an oxygen atmosphere, resulting in the formation of a layer of silicon dioxide (SiO$_2$) with a thickness of 2 $\mu$m on the silicon surface .
\item A film of 0.4 $\mu$m thick silicon nitride (Si$_3$N$_4$) was deposited onto the oxide layer by low‑pressure chemical vapor deposition (LPCVD).
\item The mask with the desired geometry was patterned on the sample by electron-beam lithography.
\item Reactive ion etching was performed through the obtained mask. During this process, the material was removed from the areas not protected by the mask down to the SiO$_2$ layer. After etching was completed, the remaining mask residues were removed by selective chemical etching.
\end{enumerate}

SEM images of the etched resonant structure are presented in Figure \ref{SEM_img}. The geometrical properties of the RW closely match those obtained from the numerical simulations. In addition to the resonant waveguides, a conventional strip waveguide was fabricated with the same height and width as the resonant one, serving as a reference.

\bmsubsection{InSe flakes preparation\label{exf}}
\vspace*{12pt}
Two InSe flakes with comparable thicknesses of $\sim$13 nm (16 monolayers) were prepared using a dry-transfer technique. The selected thickness simultaneously provides intense PL signal \cite{Borodin2024} and efficient coupling of OP exciton emission to the resonant mode of the RW \cite{Rakhlin2025}. The flakes were then placed sequentially at the center of the RW and CW structures. An optical image of the RW structure with the transferred flake is shown in Fig.~\ref{SEM_img}c. Atomic force microscopy topography confirms the surface uniformity of the flakes over the entire area, and additional $\mu$-Raman mapping across the full active region shows no qualitative or quantitative differences relative to planar InSe within the experimental uncertainty (see Supporting Information for AFM and $\mu$-Raman data).

\begin{figure}[t]
    \centering
    \includegraphics[width=0.85\linewidth]{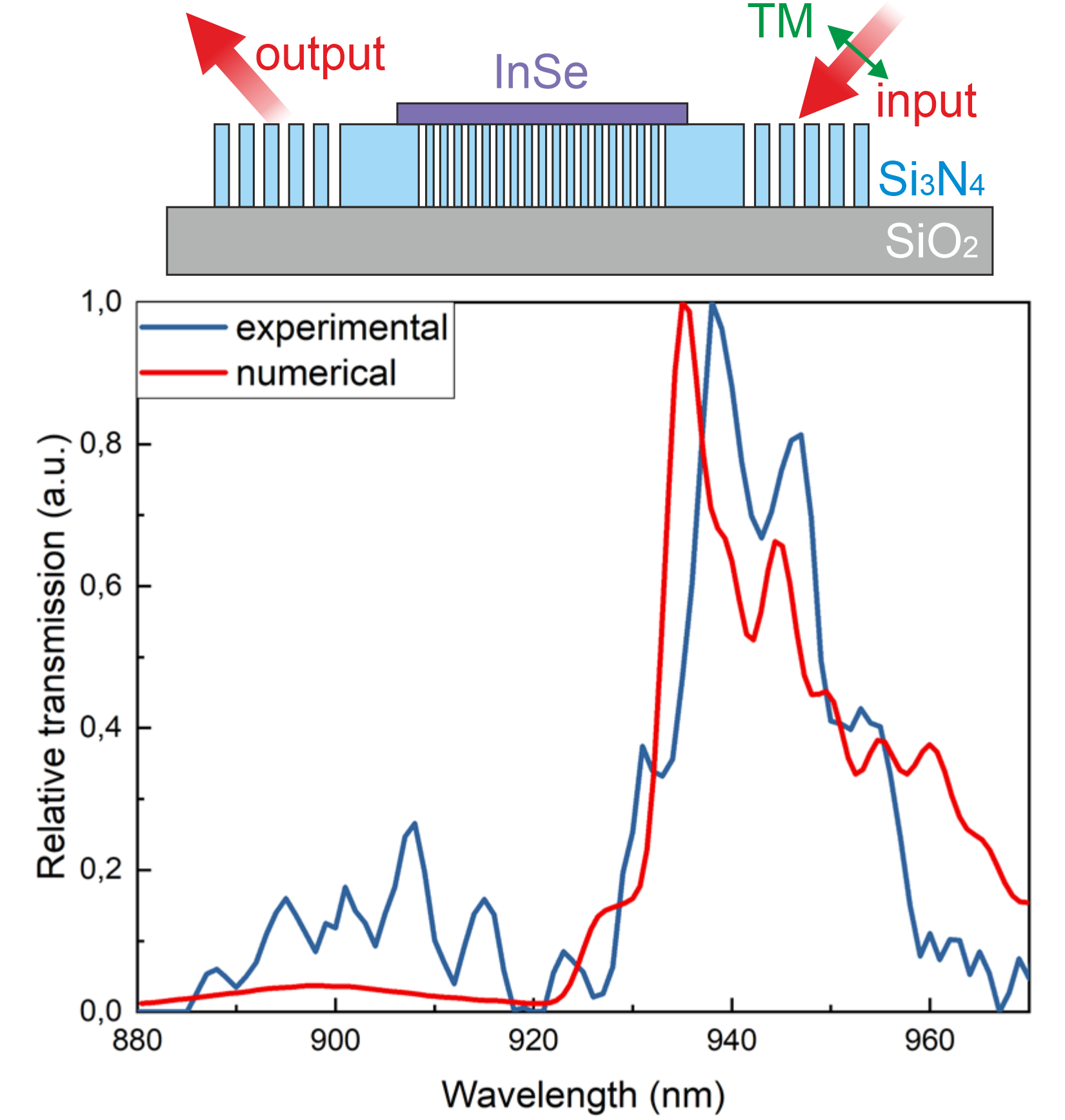}
    \caption{Relative transmission spectra of the RW structure. The sketch above the graph shows the experimental configuration for optical spectroscopy measurements.}
    \label{Rel_transm}
\end{figure}

\bmsection{Transmission spectrum measurements and simulation\label{charact}}
The sketch on top of the Figure \ref{Rel_transm} illustrates the experimental configuration. A TM-polarized laser beam was focused onto the input diffraction grating, and the radiation transmitted through the resonant structure was collected from the output grating using a high-NA objective. To obtain the relative transmittance, the power measured from the RW structure was normalized to the transmission spectrum of the reference CW. The experimentally obtained relative transmittance spectrum is shown by the blue curve in Fig. \ref{Rel_transm}, while the red curve represents the relative transmittance calculated from FDTD simulations. In the simulation, a Gaussian beam source, with  the polarization corresponding to the TM-mode of the waveguide, was placed above the input grating, and a power monitor was positioned at the output grating to record the transmitted radiation. 
The measured transmission spectrum illustrates the presence of peaks around 925~nm and 948~nm close to calculated spectral positions, where the optical coupling efficiency between excitonic dipoles in the InSe film and the resonant nanoparticles is expected to be maximized. The FDTD calculations reproduce experimental data well, minor discrepancies may be caused by fluctuation of geometrical parameters during the electron beam etching.

\begin{figure}[htbp]
    \centering
    \includegraphics[width=\linewidth]{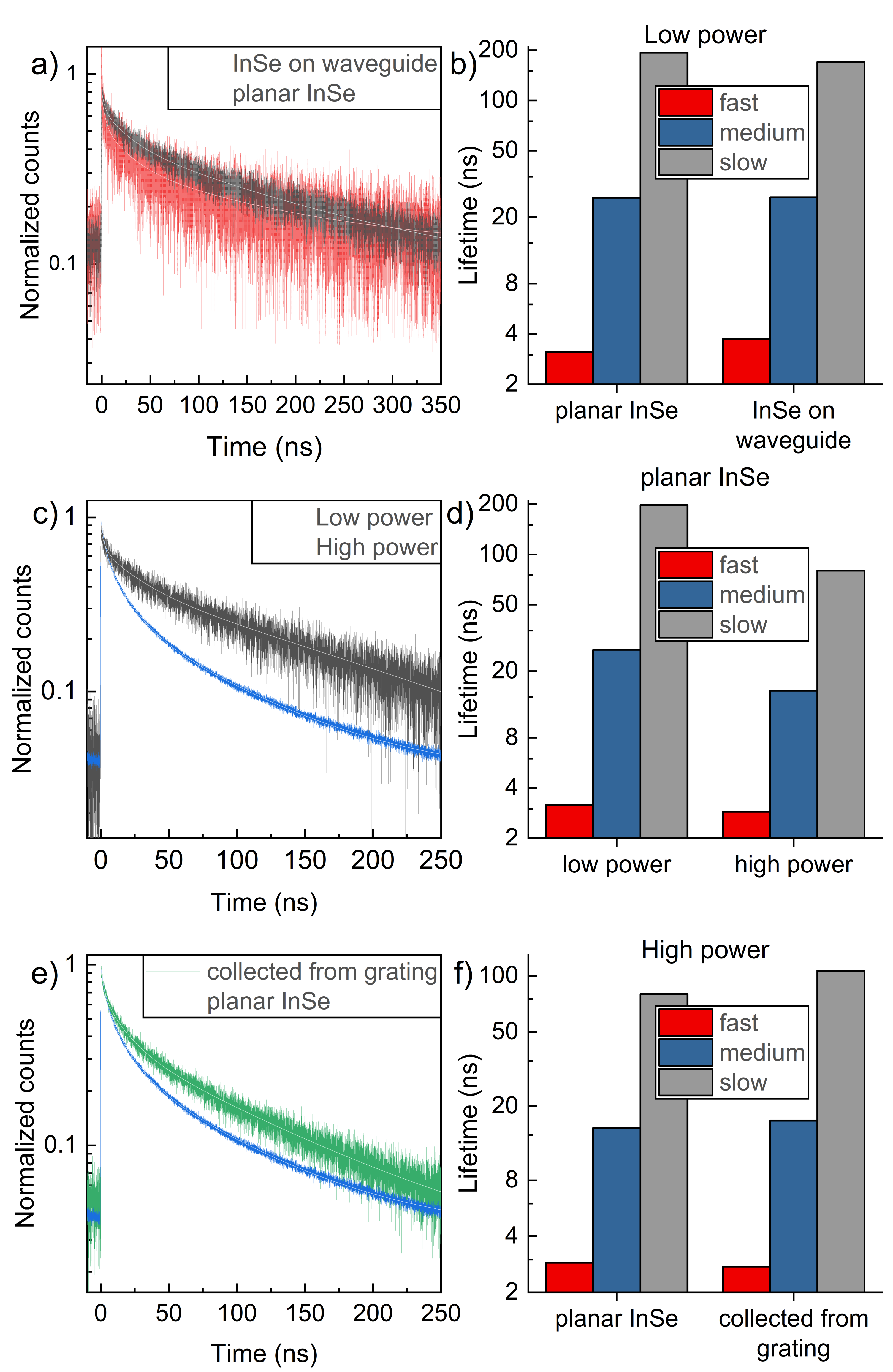}
    \caption{(a,c,e) TRPL curves obtained for the CW structure: (a) TRPL of the InSe flake at low excitation power on the substrate (gray) and the waveguide (red), (c) TRPL for InSe on the substrate at low (gray) and high (blue) excitation power, (e) TRPL for InSe luminescence collected at high excitation power from InSe on the substrate (blue) and from the grating (green). (b,d,f) Radiative lifetimes extracted from (a,c,e), respectively. Solid lines represent exponential fits.}
    \label{B4_TRPL}
\end{figure}

\bmsection{Time-resolved PL measurements}\label{TRsectionSI}
\bmsubsection{TRPL methods\label{TRmes}}
\vspace*{12pt}
The PL kinetics studies were performed at 8~K using a flow-through helium cryostat. PL excitation at 720 nm was performed by a femtosecond pulsed parametric amplifier with a pulse repetition rate of 2.6 MHz. Superconducting single-photon detectors with a time resolution of about 40 ps were used as TRPL detectors. A 920-960 nm spectral window was selected using bandpass filters.

\bmsubsection{TRPL data for the conventional waveguide\label{B4TRPL}}
\vspace*{12pt}

To verify that the observed lifetime shortening is associated with the resonant structure, we measured TRPL decay curves for the CW structure with another InSe flake of similar thickness. TRPL curves for the CW structure are presented in Figure \ref{B4_TRPL}(a,c,e), while bars in \ref{B4_TRPL}(b,d,f) show comparison of estimated lifetimes. Due to the low transmission of diffraction gratings, we increased the excitation power during the TRPL measurements of the PL signal guided from the flake to the grating. As can be seen from Figure \ref{B4_TRPL}e and \ref{B4_TRPL}f, in the case of the CW structure no lifetime reduction was observed in comparison with the planar InSe PL kinetics. During all PL measurements excitation power density was set below 500 W/cm$^2$, which allows us to exclude the saturation of the PL signal.

\bmsection{Analytical solution and parameters for the rate equations\label{SIrateq}}
\vspace*{12pt}
The system of rate equations (Eq. \ref{rateq}) can be solved analytically:
\[\begin{aligned}
    &N_{IP}(t) = \alpha\exp\left(-\left(\frac{F_p^{IP}}{\tau_{IP}}+\gamma\right)t\right) = \alpha\exp(-\Gamma_{IP} t),\\ \\
&N_{OP}(t) = A\exp\left(-\frac{F_p^{OP}}{\tau_{OP}}t\right)+B\exp\left(-\Gamma_{IP} t\right),\\ \\
&A=\frac{\gamma  \tau_{IP} \tau_{OP}-(1-\alpha) (F_p^{OP}\tau_{IP}-F_p^{IP}\tau_{OP})}{F_p^{OP}\tau_{IP}-F_p^{IP}\tau_{OP}-\gamma\tau_{OP}\tau_{IP}}.\\ \\
&B=\frac{\alpha  \gamma  \tau_{IP} \tau_{OP}}{F_p^{OP}\tau_{IP}-F_p^{IP}\tau_{OP}-\gamma\tau_{OP}\tau_{IP}}.
\end{aligned}\]
Here, $\Gamma_{IP}$ is the effective recombination rate of the IP excitons observed in TRPL measurements. For the values used at the end of Section \ref{TRSection}, we found $\Gamma_{IP}^{-1}>\tau_{OP}/P$ for $F_p^{IP,OP}>0.9$. A clear two-exponential decay requires $B>0$, which gives an upper bound $\alpha\lesssim 0.9$. In experiment this condition is easily fulfilled. Previous absorbance calculations \cite{Brotons-Gisbert2019,Brotons-gisbert2016,Mazumder2020} indicate that the initial fraction of IP excitons is much smaller: $\alpha\approx0.05$. This value is consistent with the experimentally observed relative amplitude of the middle component in the TRPL decay. The values of $F_\mathrm{p}^{IP}$ and $F_\mathrm{p}^{OP}$ were estimated from the numerical calculations shown in Figure~\ref{numcalc} by weighting the Purcell factor of each resonance by the corresponding coupling efficiency. The weighted Purcell factor, $F_\mathrm{p}^{(w)}$, was calculated as follows:
\[F_{p(w)}=\frac{\sum_iF_p^i\eta_i}{\sum_i\eta_i},\]
where $i=TE,TM$ indicates the waveguide mode, $F_p^i$ is the Purcell factor evaluated at the resonance of mode $i$ and $\eta_i$ is the coupling coefficient at the resonant wavelength for the same mode. For IP excitons, the summation is additionally performed over the two in-plane dipole orientations, \(x\) and \(y\). Using this formula, the weighted Purcell factors for OP and IP excitons were estimated to be approximately 2.16 and 2.27, respectively. To account for the difference between the idealized numerical model and the experimental structure, the weighted Purcell factors were scaled by the ratio $F_{\mathrm{p,exp}}^{OP}/F_{\mathrm{p,num}}^{OP}$, where $F_{exp}^{OP}$ is determined experimentally from the shortening of the fast component, and $F_{num}^{OP}$ is the calculated weighted Purcell factor for OP excitons.

\nocite{*}

\end{document}